\documentclass[times, 10pt,twocolumn]{article}

\usepackage{times}
\usepackage{latex8}
\usepackage{cite}
\usepackage{graphicx}
\usepackage{url}
\newcommand{\ie}{\emph{i.\,e.},\ }
\newcommand{\eg}{\emph{e.\,g.},\ }
\usepackage{url} 
\usepackage{graphicx}
\usepackage{latexsym}
\usepackage{cite}

\newtheorem{question}{RQ}

\usepackage{xspace}

\newcommand{\etal}{et\,al.\xspace}

\begin{document}

\title{Do Code Clones Matter?}

\author{Elmar Juergens, Florian Deissenboeck, Benjamin Hummel, Stefan Wagner\\
Institut f\"ur Informatik, Technische Universit\"at M\"unchen\\
Boltzmannstr. 3, 85748 Garching b.~M\"unchen, Germany\\
\{juergens,deissenb,hummelb,wagnerst\}@in.tum.de
}

\maketitle
\thispagestyle{empty}

\begin{abstract}

Code cloning is not only assumed to inflate maintenance costs but also
considered defect-prone as inconsistent changes to code duplicates can lead
to unexpected behavior. Consequently, the identification of duplicated
code, clone detection, has been a very active area of research in recent
years. Up to now, however, no substantial investigation of the consequences
of code cloning on program correctness has been carried out. To remedy this
shortcoming, this paper presents the results of a large-scale case study
that was undertaken to find out if inconsistent changes to cloned code can
indicate faults. For the analyzed commercial and open source systems we
not only found that inconsistent changes to clones are very frequent but
also identified a significant number of faults induced by such changes. The
clone detection tool used in the case study implements a novel algorithm
for the detection of inconsistent clones. It is available as open source to
enable other researchers to use it as basis for further investigations.

\end{abstract}

\vspace{-3mm}
\Section{Clones \& correctness}
\vspace{-2mm}

Research in software maintenance has shown that many programs contain a
significant amount of duplicated (cloned) code. Such cloned code is
considered harmful for two reasons: (1)~multiple, possibly unnecessary,
duplicates of code increase maintenance costs and, (2)~inconsistent changes
to cloned code can create faults and, hence, lead to incorrect program
behavior~\cite{2007_KoschkeR_survey, 2007_RoyC_Survey}. While clone
detection has been a very active area of research in recent years, up to
now, there is no thorough understanding of the degree of harmfulness of
code cloning. In fact, some researchers even started to doubt the
harmfulness of cloning at all~\cite{kapser-2006-harmful2}.

\begin{figure*}
\centering
\includegraphics[scale=.5]{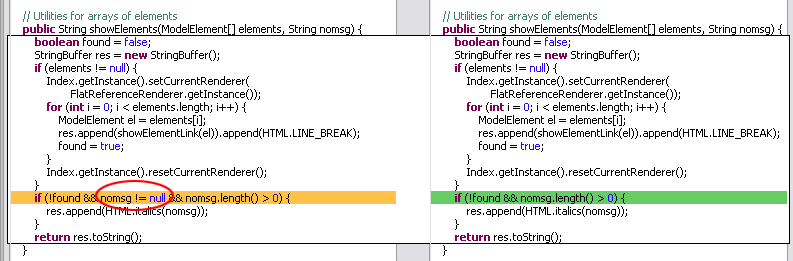} 
\caption{Missing null check on right side can cause exception (Sysiphus).}
\label{f:clone_npe}
\end{figure*}

To shed light on the situation, we investigated the effects of code cloning
on program correctness. It is important to understand, that clones do not
directly cause faults but inconsistent changes to clones can lead to
unexpected program behavior. A particularly dangerous type of change to
cloned code is the \emph{inconsistent bug fix}. If a fault was found in
cloned code but not fixed in \emph{all} clone instances, the system is
likely to still exhibit the incorrect behavior. To illustrate this,
Fig.~\ref{f:clone_npe} shows an example, where a missing null-check was
retrofitted in only one clone instance.

This paper presents the results of a large-scale case study that was undertaken
to find out (1) if clones are changed inconsistently, (2) if these
inconsistencies are introduced intentionally and, (3) if unintentional
inconsistencies can represent faults. In this case study we analyzed three
commercial systems written in C\#, one written in Cobol and one open-source
system written in Java. To conduct the study we developed a novel detection
algorithm that enables us to detect inconsistent clones. We \emph{manually}
inspected about 900 clone groups to handle the inevitable false positives and
discussed each of the over 700 inconsistent clone groups with the
\emph{developers} of the respective systems to determine if the inconsistencies
are intentional and if they represent faults. Altogether, around 1800
individual clone group assessments were manually performed in the course of the
case study. The study lead to the identification of 107 faults that have been
confirmed by the systems' developers.

\vspace{-.7em}
\paragraph{Research Problem} Although most previous work
agrees that code cloning poses a problem for software maintenance, ``there
is little information available concerning the impacts of code clones on
software quality''~\cite{2007_RoyC_Survey}. As the  consequences of code
cloning on program correctness, in particular, are not fully understood
today, it remains unclear how \emph{harmful} code clones really are.
We consider the absence of a thorough understanding of code cloning
precarious for software engineering research, education and practice.

\vspace{-.7em}
\paragraph{Contribution} The contribution of this paper is twofold. First,
we extend the existing empirical knowledge by a case study that demonstrates
that clones get changed inconsistently and that such changes can represent
faults. Second, we present a novel suffix-tree based algorithm for the
detection of inconsistent clones. In contrast to other algorithms for the
detection of inconsistent clones, our tool suite is made available for other
researchers as open source.

\Section{Terms and definitions}

The literature provides a wide variety of different definitions of
clones and clone related terms~\cite{2007_KoschkeR_survey,
2007_RoyC_Survey}. To avoid ambiguity, we describe the terms as used
in this paper.

\emph{Code} is interpreted as a sequence of \emph{units}, which for example
could be characters, normalized statements, or lines. The reason to allow
normalization of units at this stage, is that often pieces of code are
considered equal even despite differences in comments or naming, which can be
leveled by the normalization.  An \emph{exact clone} is then a (consecutive)
substring of the code that appears at least twice in the (normalized) code.
Thus our definition of a clone is purely syntactical, but catches exactly the
idea of \emph{copy\&paste}, while allowing simple changes, such as renaming, due
to normalization.  An \emph{exact clone group} is a set of at least two
exact clones that appear at different positions.

To capture the notion of non-identical clones, we roughly follow the
definitions of a gapped or type 3 clone given
in~\cite{2007_KoschkeR_survey,2007_RoyC_Survey}.  A substring $s$ of the code
is called an \emph{inconsistent clone}, if there is another substring $t$ of the
code such that their edit distance is below a given threshold and that $t$ has no
significant overlap with $s$. The edit distance is a metric
that counts the number of edit operations (insertion, removal, or change of a
single unit) needed to transform one sequence into the other.  Obviously, this
definition is slightly vague, as it depends on the threshold chosen and the
meaning of a ``significant overlap''. However, it captures our intuitive
understanding of an inconsistent clone as used in this paper. Examples are
shown in Figs.~\ref{f:clone_npe} and~\ref{f:clone_ui}. By \emph{clone} we
denote both exact and inconsistent clones.

A \emph{clone group} can be viewed as a \emph{connected} graph, where each
node is a substring, and edges are drawn between substrings that are
clones of each other. If at least one pair of inconsistent clones is
in the group, it is called an \emph{inconsistent clone group}.  We could
also have required all clones in a clone group to be clones of each
other, but often these slightly larger clone groups created by our
definition reveal interesting relationships in the code.

For a thorough discussion of the consequences of inconsistent clones, we
define that a \emph{failure} is an incorrect output of a software visible
to the user and that a \emph{fault} is the cause of a potential failure
inside the code. \emph{Defects} are the superset of faults and failures.

\Section{Related work}

A substantial amount of research has been dedicated to code cloning in
recent years. The detailed surveys by Koschke
\cite{2007_KoschkeR_survey} or Roy and Cordy \cite{2007_RoyC_Survey}
provide a comprehensive overview of existing work. Since this paper
targets consequences of cloning and detection of inconsistent clones,
we detail existing work in these areas.

\subsection{Consequences of cloning}

Indication for harmfulness of cloning for maintainability or correctness is
given by several researchers. Lague \etal \cite{1997_Lague}, report inconsistent
evolution of a substantial amount of clones in an industrial telecommunication
system. Monden \etal \cite{2002_Monden_clonging_impact_on_maintainability}
report a higher revision number for files with clones than for files without in
a 20 year old legacy system, possibly indicating lower maintainability. In
\cite{2005_Kim}, Kim \etal report that a substantial
amount of changes to code clones occur in a coupled fashion, indicating
additional maintenance effort due to multiple change locations.

Li \etal \cite{2006_Li_cpminer} present an approach to detect bugs based on
inconsistent renaming of identifiers between clones. Jiang, Su and Chiu
\cite{jiang-2007-clonebugs} analyze different contexts of clones, such as
missing \emph{if} statements. Both papers report the successful discovery of
bugs in released software. In \cite{2007_Aversano_clone_maintenance} and
\cite{2007_Tibor_clone_smells}, individual cases of bugs or inconsistent bug
fixes discovered by analysis of clone evolution are reported for open source
software.

In contrast, doubt that consequences of cloning are unambiguously harmful is
raised by several recent research results. Krinke
\cite{2007_Krinke_consistent_and_inconsistent_changes} reports that only half
the clones in several open source systems evolved consistently and that only a
small fraction of inconsistent clones becomes consistent again through later
changes, potentially indicating a larger degree of independence of clones than
hitherto believed. Geiger \etal \cite{Geiger06relationof} report that a
relation between change couplings and code clones could, contrary to
expectations, not be statistically verified. Lozano and Wermelinger
\cite{2008_Lozano_clones_changeability} report that no systematic relationship
between code cloning and changeability could be established.

The effect of cloning on maintainability and correctness is thus not clear.
Furthermore, the above listed publications suffer from one or more shortcomings
that limit the transferability of the reported findings.

\begin{itemize}
  \item Instead of manual inspection of the actual inconsistent clones to
  evaluate consequences for maintenance and correctness, indirect
  measures\footnote{Examples are change coupling or the ratio between
  consistent and inconsistent evolution of clones} are used \cite{1997_Lague,
  2002_Monden_clonging_impact_on_maintainability,
  2007_Krinke_consistent_and_inconsistent_changes,
  2007_Aversano_clone_maintenance, 2008_Lozano_clones_changeability,
  Geiger06relationof}. Such approaches are inherently inaccurate and can easily
  lead to misleading results. For example, unintentional differences and
  faults, while unknown to developers, exhibit the same evolution pattern as
  intentional independent evolution and are thus prone to misclassification.

  \item The analyzed systems are too small to be representative \cite{2005_Kim} or omit analysis of industrial software
  \cite{2007_Tibor_clone_smells, 2005_Kim,
  2007_Krinke_consistent_and_inconsistent_changes,
  2007_Aversano_clone_maintenance, 2008_Lozano_clones_changeability,
  Geiger06relationof}.

  \item The analyses specifically focus on faults introduced during creation
  \cite{2006_Li_cpminer, jiang-2007-clonebugs} or evolution
  \cite{2007_Tibor_clone_smells} of clones, inhibiting quantification of inconsistencies in general.
\end{itemize}

Additional empirical research outside these limitations is required to
better understand consequences of cloning \cite{2007_KoschkeR_survey, 2007_RoyC_Survey}, as presented in this paper:
Developer rating of the actual inconsistent clones has been performed, the
study objects are both open source and industrial systems and inconsistencies
have been analyzed independently of their mode of creation.

\subsection{Detection of inconsistent clones}

We classify existing approaches according to the program representation on
which they operate.

\noindent\textbf{Text~} Normalized code fragments are compared textually in a
pairwise fashion \cite{2008_Roy_nicad}. A similarity threshold governs
whether text fragments are considered as clones.

\noindent\textbf{Token~} Ueda \etal \cite{2002_Uedaa_gapped_clones} propose
post-processing of the results of a token-based detection of exact clones.
Essentially, neighboring exact clones are composed into inconsistent clones.
In \cite{2006_Li_cpminer}, Li \etal present the tool CP-Miner, which searches
for similar basic blocks using frequent subsequence mining and then combines
basic block clones into larger clones.

\noindent\textbf{Abstract Syntax Tree~} Baxter \etal \cite{1998_Baxter} hash subtrees
into buckets and perform pairwise comparison of subtrees in the same bucket.
Jiang \etal \cite{2007_JiangL_decard} propose the generation of
characteristic vectors for subtrees. Instead of pairwise comparison, they
employ locality sensitive hashing for vector clustering, allowing for better
scalability than \cite{1998_Baxter}. In
\cite{2007_Evans_clone_detection_structural_abstraction}, tree patterns that
provide structural abstraction of subtrees are generated to identify cloned
code.

\noindent\textbf{Program Dependence Graph~} Krinke \cite{2001_Krinke} proposes a search
algorithm for similar subgraph identification. Komondoor and Horwitz
\cite{2001_Komondoor} propose slicing to identify isomorphic PDG subgraphs.
Gabel, Jiang and Su \cite{gabel-2008-semanticclones} use a modified slicing
approach to reduce the graph isomorphism problem to tree similarity.

The existing approaches provided valuable inspiration for the algorithm
presented in this paper. However, none of them was applicable to our case
study, for one or more of the following reasons.

\begin{itemize}
  \item Tree \cite{1998_Baxter, 2007_JiangL_decard,
  2007_Evans_clone_detection_structural_abstraction} and graph
  \cite{2001_Krinke, 2001_Komondoor, gabel-2008-semanticclones} based
  approaches require the availability of suitable context free grammars for AST
  or PDG construction. While feasible for modern languages such as Java, this
  poses a severe problem for legacy languages such as Cobol or PL/I, where
  suitable grammars are not available. Parsing such languages still represents
  a significant challenge \cite{LV01-SPE, 1999_Ducasse}.

  \item Due to the information loss incurred by the reduction of variable size
  code fragments to finite-size numbers or vectors, the edit distance between
  inconsistent clones cannot be precisely controlled in feature vector
  \cite{2007_JiangL_decard} and hashing based \cite{1998_Baxter} approaches.

  \item Idiosyncrasies of some approaches threaten recall. In
  \cite{2002_Uedaa_gapped_clones}, inconsistent clones cannot be detected if
  their  constituent exact clones are not long enough. In
  \cite{gabel-2008-semanticclones}, inconsistencies might not be detected if
  they add data or control dependencies, as noted by the authors.

  \item Scalability to industrial-size software of some approaches has been
  shown to be infeasible \cite{2001_Krinke, 2001_Komondoor} or is at least
  still unclear \cite{2007_Evans_clone_detection_structural_abstraction,
  2008_Roy_nicad}.

  \item For most approaches, implementations are not publicly available.
\end{itemize}

In contrast, the approach presented in this paper supports both
modern and legacy languages including Cobol and PL/I, allows for precise
control of similarity in terms of edit distance on program statements, is
sufficiently scalable to analyze industrial-size projects in reasonable time
and is available for use by others as open source software.

An approach similar to \cite{2002_Uedaa_gapped_clones} for bug detection has
been outlined by the authors of this paper in \cite{2008_Juergens_Teso}. In
contrast to this work, it does not use a suffix tree based algorithm and no
empirical study was performed.

\Section{Detecting inconsistent clones}
\label{sec:approach}

This section explains the approach used for detecting inconsistent
clones in large amounts of code. Our approach works on the token
level, which usually is sufficient for finding copy-pasted code, while
at the same time being efficient. The algorithm works by
constructing a suffix tree of the code and then for each possible suffix
an approximate search based on the edit distance in this tree is performed.

\begin{figure}
\centering
\includegraphics[width=.9\columnwidth]{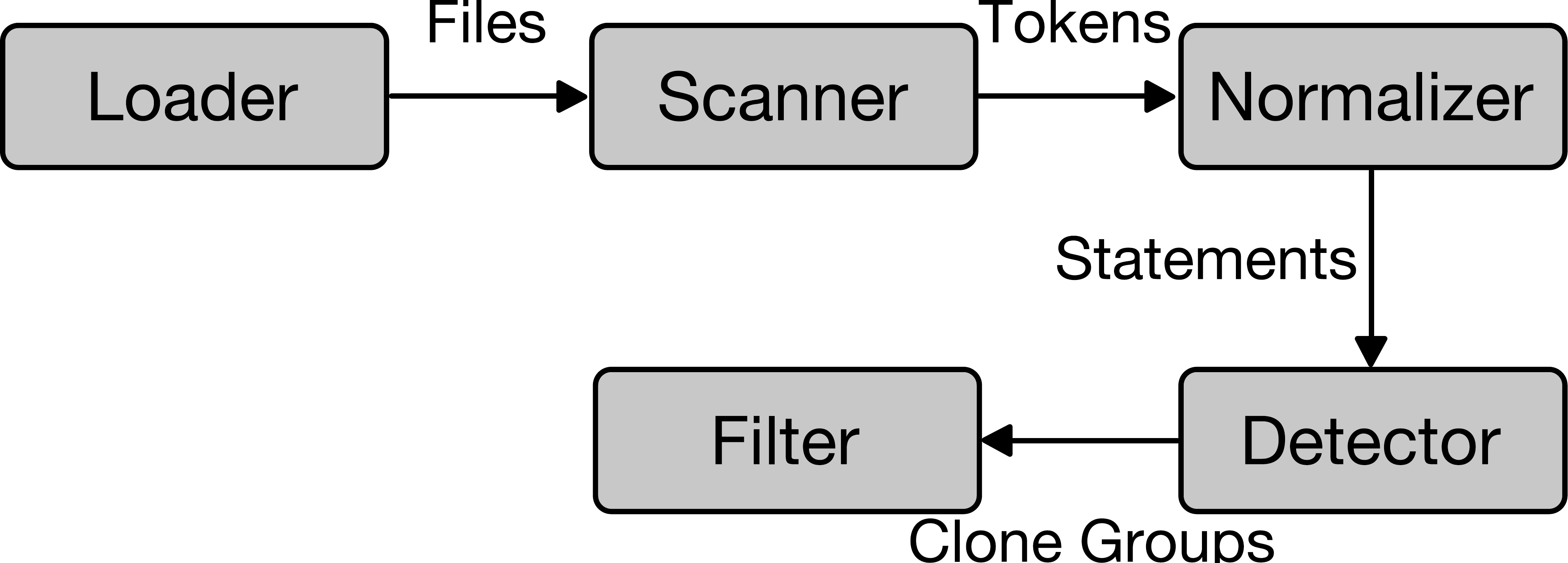}
\caption{The clone detection pipeline used}
\label{f:pipeline}
\end{figure}

Our clone detector is organized as a pipeline, which is sketched in
Figure~\ref{f:pipeline}. The files under analysis are loaded and then
fragmented by the \emph{scanner}, yielding a stream of tokens, which is
filtered to exclude comments and generated code (recognized by user provided
patterns).  From the token stream, which consist of single keywords,
identifiers, operators, and so on, the \emph{normalizer} reassembles
statements. This stage performs normalization, such that differences in
identifier names or constant values are not relevant when comparing statements.
The sequence formed by those statements is then fed into our clone detection
algorithm, which finds and reports clone groups in this stream. Finally, clone
groups are post-processed and uninteresting ones are filtered out. We
outline the detection steps in more detail in the following subsections.

\SubSection{Preprocessing and normalization}

As stated before, the code is read and split into tokens using a scanner. An
important task during preprocessing is normalization, which creates
statements from the scanner's tokens. This is used as it allows better
tailoring of normalization and to avoid clones starting or ending within
statements.  The used normalization eliminates differences in naming of identifiers
and values of constants or literals, but does not, for example, change
operation order.

Further tasks of the preprocessing phase are the removal of comments
or generated code, which is either already excluded at the file level
or on the token stream based on certain patterns that recognize
sections of generated code.

\SubSection{Detection algorithm}

The task of the detection algorithm is to find clones in the stream of
units provided by the normalizer. Stated differently, we want to
find common substrings in the sequence formed by all units of the
stream, where common substrings are not required to be exactly
identical (after normalization), but may have an edit distance bounded
by some threshold. This problem is related to the approximate string
matching problem~\cite{jokinen-1991-approx,ukkonen-93-approximate},
which is also investigated extensively in
bioinformatics~\cite{taeubig-2007-diss}. The main difference is that
we are not interested in finding an approximation of only a single
given word in the string, but rather are looking for \emph{all}
substrings approximately occurring more than once in the entire
sequence.

A sketch of our detection algorithm is shown in
Figs.~\ref{fig:algo_detect} and~\ref{fig:algo_search}.  The
algorithm is an edit distance based traversal of a suffix
tree of our input sequence. A suffix tree over a sequence $s$ is a
tree with edges labeled by words such that exactly all suffixes of $s$
are found by traversing the tree from the root node to a leaf and
concatenating the words on the edges encountered. Such a suffix tree
can be constructed in linear time by the well-known online algorithm
by Ukkonen~\cite{ukkonen-1995-suffix}. Using this suffix tree, we
start a search for clones at every possible index.

Searching for clones is performed by the procedure \emph{search}
which recursively traverses the suffix tree.  The first two parameters
to this function are the sequence $s$ we are working on and the
position \emph{start} where the search was started, which is required
when reporting a clone. The parameter $j$ (which is the same as
\emph{start} in the first call of \emph{search}) marks the current end
of the substring under inspection. To prolong this substring, the
substring starting at $j$ is compared to the word $w$ being next in
the suffix tree, which is the edge leading \emph{to} the current node
$v$ (for the root node we just use the empty string). For this
comparison an edit distance of at most $e$ operations (fifth
parameter) is allowed. For the first call of \emph{search}, $e$ is the
edit distance maximally allowed for a clone. If the remaining
edit operations are not enough to \emph{match} the entire edge word
$w$ (else case), we report the clone as far as we found it, otherwise
the traversal of the tree continues recursively, increasing the length
($j - $\emph{start}) of the current substring and reducing the
number $e$ of edit operations available by the amount of operations
already spent in this step.

\begin{figure}[htb]
  \small
  \begin{tabular}{rl}
    \hline
    & \textbf{proc} detect ($s$, $e$) \\
    & \textbf{Input: } String $s = (s_0, \ldots, s_n)$,
       max edit distance $e$ \vspace{2mm} \\
    1 & Construct suffix tree $T$ from $s$ \\
    2 & \textbf{for each} $i \in \{1, \ldots, n\}$ \textbf{do} \\
    3 & \quad search ($s$, $i$, $i$, root($T$), $e$) \\
    \hline
  \end{tabular}
  \caption{Outline of approximate clone detection algorithm}
  \label{fig:algo_detect}
\end{figure}

\begin{figure}[htb]
  \small
  \begin{tabular}{rl}
    \hline
    & \textbf{proc} search ($s$, start, $j$, $v$, $e$) \\
    & \textbf{Input: } String $s = (s_0, \ldots, s_n)$, \\
    & \quad start index of current search, current search index $j$, \\
    & \quad node $v$ of suffix tree over $s$, max edit distance $e$ \vspace{2mm} \\

    1 & Let $(w_1, \ldots, w_m)$ be the word along the edge
        leading to $v$ \\
    2 & Calculate the maximal length $l \le m$, such that \\
      & there is a $k \ge j$ where the edit distance $e'$ between \\
      & $(w_1, \ldots, w_l)$ and $(s_j, \ldots, s_k)$ is at most $e$ \\
    3 & \textbf{if} $l = m$ \textbf{then} \\
    4 & \quad \textbf{for each} child node $u$ of $v$ \textbf{do} \\
    5 & \quad \quad search ($s$, start, $k+m$, $u$, $e - e'$) \\
    6 & \textbf{else if} $k - \textrm{start} \ge$ minimal clone length \textbf{then} \\
    7 & \quad report substring from \emph{start} to $k$ of $s$ as clone \\
    \hline
  \end{tabular}
  \caption{Search routine of the approximate clone detection
  algorithm}
  \label{fig:algo_search}
\end{figure}

To actually make this algorithm work and its results usable, some
details have to be fleshed out. For the computation of the longest
edit distance match we are using the simple dynamic programming
algorithm found in algorithm textbooks. While this is easy to
implement, it requires quadratic time and space\footnote{Actually the
algorithm can be implemented using only linear space, but preserving
the full calculation matrix allows us some simplifications.}. To make
this step work efficiently we look at most at the first 1000
statements of the word $w$. As long as the word on the suffix tree
edge is shorter, this is not a problem. In case there is a clone of
more than 1000 statements, we would find it in chunks of 1000. We
considered this to be tolerable for practical purposes.  As each
suffix we are running the search on will of course be part of the
tree, we also have to make sure that no self matches are reported.

When running the algorithm as it is, the results are often not as
expected because the search tries to match as many statements as
possible. However, allowing for edit operations right at the beginning or at
the end of a clone is not helpful, as then every exact clone can
be prolonged into an inconsistent clone. Thus in the search we enforce
the first few statements (how many is parameterized) to match exactly.
(This also speeds up the search, as we can choose the correct child
node at the root of the suffix tree in one step without looking at all
children.) The last statements are also not allowed to differ, which
is checked for and corrected just before reporting a clone.

Including all of these optimizations, the algorithm can miss a clone
either due to the thresholds (either too short or too many
inconsistencies), or if it is covered by other clones. The later case
is important, as each substring of a clone of course is a clone again
and we usually do not want these to be reported.

\SubSection{Post-processing and filtering}

During and after detection, the clone groups that are reported are subject to
filtering. Filtering is usually performed as early as possible, so no memory is
wasted with storing clone groups that are not considered relevant. Using these
filters, we discard clone groups whose clones overlap with each other and
groups whose clones are contained in other clone groups. Additionally, we
enforce not only an absolute limit on the number of inconsistencies, but also a
relative one, \ie we filter clone groups where the number of inconsistencies in
the clones relative to the clone's length exceeds a certain amount. Moreover,
we merge clone groups which share a common clone. While this leads to clone
groups with non related clones (as our definition of an inconsistent clone is
not transitive), for practical purposes it is preferred to know of these
indirect relationships, too.

\SubSection{Tool support}
\label{sec:tools}

To be able to experiment with the detection of inconsistent clones,
our algorithms and filters have been implemented as part of
CloneDetective\footnote{Available as Open Source
\url{http://www.clonedetective.org}}
\cite{Juergens2009_clonedetective} which is based on
ConQAT~\cite{2008_deissenboeckf_quality_control}. The result is a
highly configurable and extensible platform for clone detection on the
syntactic level. As our cloning pipeline could reuse a major portion
of the CloneDetective code, we consider such an open platform
essential for future experiments, as it allows researchers to focus on
individual parts of the pipeline. CloneDetective also offers a
front-end to visualize and assess the clones found, and thus supports
the rapid review of a large number of clone groups.

\SubSection{Scalability and performance}

\begin{figure}
\centering
\includegraphics[width=\columnwidth]{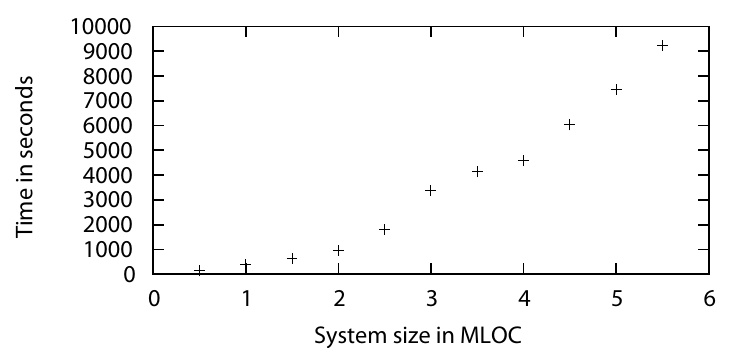}
\caption{Runtime of inconsistent clone detection on Eclipse source}
\label{f:performance}
\end{figure}

Due to the many implementation details, the worst case complexity is hard to
analyze. Additionally, for practical purposes, the more complicated average
complexity would be more adequate.  Thus, and to assess the performance of the
entire pipeline we executed the detector on the source code of
Eclipse\footnote{Core of Eclipse Europa release 3.3}, limiting detection to a
certain amount of code. Our results on an Intel Core 2 Duo 2.4 GHz running Java
in a single thread with 3.5 GB of RAM are shown in Figure~\ref{f:performance}.
The settings are the same as for the main study (min clone length of 10, max
edit distance of 5). It is capable to
handle the 5.6 MLOC of Eclipse in about 3 hours, which is fast enough to be
executed within a nightly build.

\Section{Study description}

In order to gain a solid insight into the effects of inconsistent clones,
we use a study design with 5 objects and
3 research questions that guide the investigation.

\SubSection{Study objects}

We chose 2 companies and 1 open source project as sources of software systems.
This resulted in 5 analyzed projects in total. We chose systems written in
different languages, by different teams in different companies and with
different functionalities to increase the transferability of the study
results. These objects included 3 systems written in C\#, a Java system as well
as a long-lived Cobol system. All these systems are already in production. For
non-disclosure reasons we gave the commercial systems names from A to D. An
overview is shown in Table~\ref{tab:systems}.

\paragraph{Munich Re Group} The Munich Re Group is one of the largest
re-insurance companies in the world and employs more than 37,000 people in over
50 locations. For their insurance business, they develop a variety of
individual supporting software systems. In our study, we analyzed the systems
A, B and C, all written in C\#. They were each developed by different
organizations and provide substantially different functionality, ranging from
damage prediction, over pharmaceutical risk management to credit and company
structure administration. The systems support between 10 and 150 expert users
each.

\paragraph{LV 1871} The Lebensversicherung von 1871 a.G. (LV~1871) is a
Munich-based life-insurance company. The LV 1871 develops and maintains several
custom software systems for mainframes and PCs. In this study, we analyze a
mainframe-based contract management system mostly written in Cobol (System D)
employed by about 150 users.

\paragraph{Sysiphus} The open source system
\emph{Sysiphus}\footnote{\url{http://sysiphus.in.tum.de/}} is developed at the
Technische Universit\"at M\"unchen (TUM) but none of the authors of this paper have
been involved in the development. It constitutes a collaboration environment
for distributed software development projects. The inclusion of an open source
system is motivated by the fact that, as the clone detection tool is also freely
available, the results can be externally
replicated\footnote{\url{http://wwwbroy.in.tum.de/~ccsm/icse09/}}. This is not
possible with the detailed confidential results of the commercial systems.

\begin{table}[htb]
\caption{Summary of the analyzed systems \label{tab:systems}}
\begin{center}
{\small
\begin{tabular}{|l||l|l|r|r|}
\hline
\textbf{System} & Organization & Language & Age & Size \\
& & & (years) & (kLOC) \\
\hline
\textbf{A} & Munich Re & C\# & 6 & 317 \\
\textbf{B} & Munich Re & C\# & 4 & 454 \\
\textbf{C} & Munich Re & C\# & 2 & 495 \\
\textbf{D} & LV 1871 & Cobol & 17 & 197 \\
\textbf{Sysiphus} & TUM & Java & 8 & 281 \\
\hline
\end{tabular}}
\end{center}
\end{table}

\SubSection{Research questions}

The underlying problem that we analyze are clones and
especially their inconsistencies. In order to investigate this question,
we answer the following 3 more detailed research questions.

\begin{question}
Are clones changed inconsistently?
\end{question}
The first question we need to answer is whether inconsistent clones appear
at all in real-world systems. This not only means whether we can find them
at all but also whether they constitute a significant part of the total
clones of a system. It does not make sense to analyze inconsistent
clones if they are a rare phenomenon.

\begin{question}
Are inconsistent clones created unintentionally?
\end{question}
Having established that there are inconsistent clones in real systems,
we need to analyze whether these inconsistent clones have been created
intentionally or not. It can obviously be sensible to change a clone
so that it becomes inconsistent to its counterparts because it has to
conform to different requirements. However, the important difference
is whether the developer is aware of the other clones, i.e.\ whether
the inconsistency is intentional.

\begin{question}
Can inconsistent clones be indicators for faults in real systems?
\end{question}
After establishing these prerequisites, we can determine whether the
inconsistent clones are actually indicators for faults in real systems.
If there are inconsistent clones that have not been created because of different
requirements, this implies that at least one of these clones does not conform to
the requirements. Hence, it constitutes a fault.

\SubSection{Study design}

\begin{figure}
\begin{center}
\includegraphics[width=0.8\linewidth]{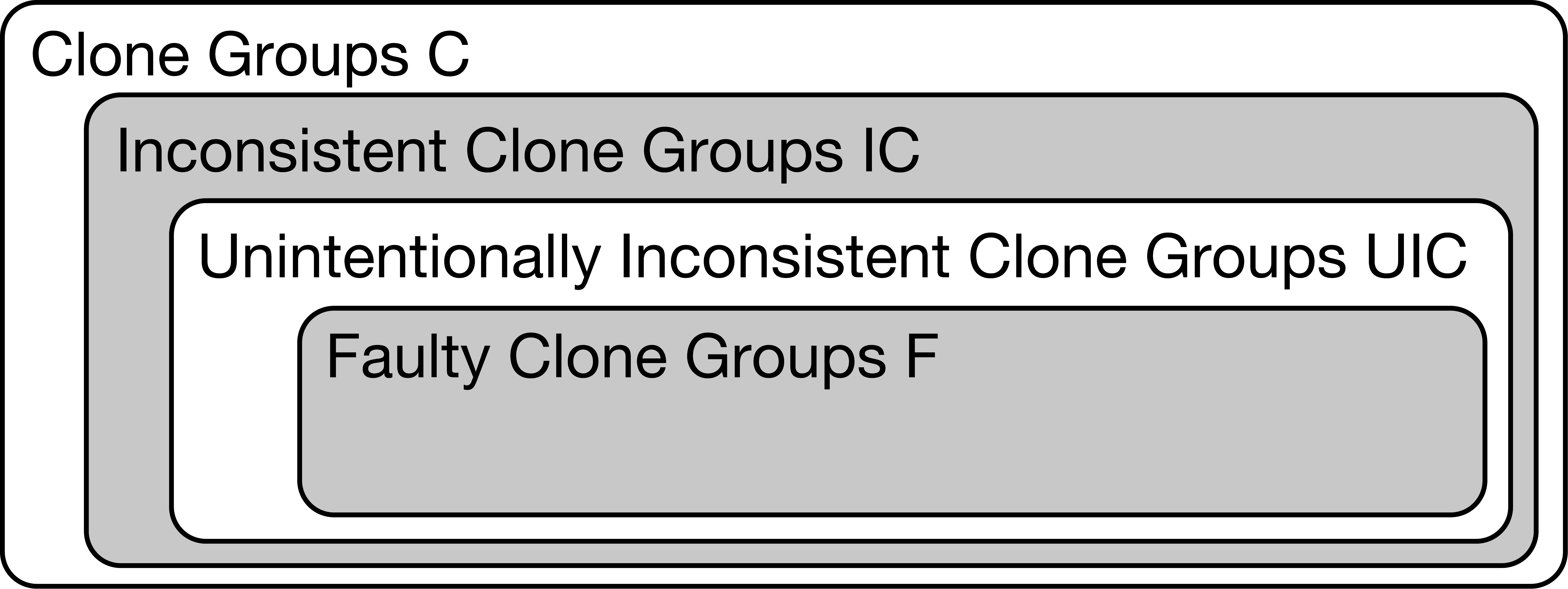}
\caption{Clone Group Sets\label{fig:clone_sets}}
\end{center}
\end{figure}

We answer the research questions with the following study design. In the study
we analyze sets of clone groups as shown in Fig.~\ref{fig:clone_sets}. The
outermost set are all clone groups $C$ in a system, \emph{IC} denotes the set
of inconsistent clone groups, and \emph{UIC} the unintentionally inconsistent
clone groups. The subset $F$ of \emph{UIC} consists of those unintentionally
inconsistent clone groups that indicate a fault in the program. Please note
that we do not distinguish between \emph{created} and \emph{evolved}
inconsistent clones as for the question of faultiness it does not matter when
the inconsistencies have been introduced.

We use these different clone group sets to design the study that answers our
research questions. The independent variables in the study are development
team, programming language, functional domain, age and size. The dependent
variables for the research questions are explained below.
RQ 1 investigates the existence of inconsistent clones in realistic systems.
Hence, we need to analyze the size of set \emph{IC} with respect to the size of
set $C$. We apply our inconsistent clone analysis approach to all the
systems, perform manual assessment of the detected clones to eliminate false
positives and calculate the \emph{inconsistent clone ratio} $|\textit{IC}| /
|C|$.

For RQ 2, whether clones are created unintentionally, we then compare the
size of the sets \emph{UIC} and \emph{IC}. The sets are established by
showing each identified inconsistent clone to developers of the system
and asking them to rate them as intentional or unintentional.
This gives us the
\emph{unintentionally inconsistent clone ratio} $|\textit{UIC}| /
|\textit{IC}|$.
The most important question we aim to answer is whether inconsistent
clones indicate faults (RQ 3). Hence, we are interested in the size of
set $F$ in relation to the size of \emph{IC}. The set $F$ is again determined by
asking developers of the respective system. Their expert opinion classifies
the clones in faulty and non-faulty. We only analyze unintentionally
inconsistent clones for faults. Our \emph{faulty inconsistent clone ratio}
$|\textit{F}| / |\textit{IC}|$ is thus a lower bound, as potential faults in
intentionally inconsistent clones are not considered.

Using this, we are already able to roughly find the answer to RQ 3. As
this is our main result from the study, we transform it into a
hypothesis. We need to make sure that the fault density in the
inconsistencies is higher than in
randomly picked lines of source code.
This leads to the hypothesis $H$:

\textit{The fault density in the inconsistencies
is higher than the average fault density.}

As we
do not know the actual fault densities of the analyzed systems, we need to
resort to average values. The span of available numbers is large because
of the high variation in software systems. Endres and Rombach \cite{endres03}
give 0.1--50 faults per kLOC as a typical range. For the fault density in the
inconsistencies, we use the number of faults divided by the logical lines
of code of the inconsistencies. We refrain from testing the hypothesis
statistically because of the low number of data points as well as the
large range of typical defect densities.

\begin{table*}[htb]
\caption{Summary of the study results \label{tab:results}}
\begin{center}
{\small
\begin{tabular}{|l||c|c|c|c|c||c|c|}
\hline
Project & \multicolumn{1}{c|}{\textbf{A}} & \multicolumn{1}{c|}{\textbf{B}} & \multicolumn{1}{c|}{\textbf{C}} & \multicolumn{1}{c|}{\textbf{D}} & \multicolumn{1}{c||}{\textbf{Sysiphus}} & \multicolumn{1}{c|}{\textbf{Sum}} & {\textbf{Mean}} \\
\hline
Precision exact clone groups        & 0.88 & 1.00 & 0.96 & 1.00 & 0.98 & --- & 0.96 \\
Precision inconsistent clone groups & 0.61 & 0.86 & 0.80 & 1.00 & 0.87 & --- & 0.83 \\
\hline
Clone groups $|C|$ & 286 & 160 & 326 & 352 & 303 & 1427 & --- \\
Inconsistent clone groups $|IC|$ & 159 & 89 & 179 & 151 & 146 & 724 & --- \\
Unintentionally inconsistent clone groups $|UIC|$ & 51 & 29 & 66 & 15 & 42 & 203 & --- \\
Faulty clone groups $|F|$ & 19 & 18 & 42 & 5 & 23 & 107 & --- \\
\hline
RQ 1 $|IC|/|C|$ & 0.56 & 0.56 & 0.55 & 0.43 & 0.48 & --- & 0.52 \\
RQ 2 $|UIC| / |IC|$ & 0.32 & 0.33 & 0.37 & 0.10 & 0.29 & --- & 0.28 \\
RQ 3 $|F| / |IC|$ & 0.12 & 0.20 & 0.23 & 0.03 & 0.16 & --- & 0.15 \\
\hline
Faulty in UIC $|F|/|UIC|$ & 0.37 & 0.62 & 0.64 & 0.33 & 0.55 & --- & 0.50 \\
\hline
Inconsistent logical lines & 442 & 197 & 797 & 1476 & 459 & 3371 & --- \\
Fault density in kLOC$^{-1}$ & 43 & 91.4 & 52.7 & 3.4 & 50.1 & --- & 48.1 \\
\hline
\end{tabular}}
\end{center}
\end{table*}

\SubSection{Procedure}

The treatment we used on the objects was the approach to detect inconsistent
clones as described in section \ref{sec:approach}. For all systems, the
detection was executed by the researcher to identify consistent and
inconsistent clone candidates. On an 1.7 GHz notebook, the detection took
between one and two minutes for each system. The detection was configured to
not cross method boundaries, since experiments showed that inconsistent clones
that cross method boundaries in many cases did not capture semantically
meaningful concepts. This is also noted for exact clones in
\cite{2006_Koschke_syntax_suffix_trees} and is even more pronounced for
inconsistent clones. Since in Cobol sections in the procedural division are the
counterpart of Java or C\# methods, clone detection for Cobol was limited to
these.

For the C\# and Java systems, the algorithm was parameterized to use
10 statements as minimal clone length, a maximum edit distance of 5, a
maximal inconsistency ratio (\ie the ratio of edit distance and
clone length) of 0.2 and the constraint that the first 2 statements of
two clones need to be equal. Due to the verbosity of
Cobol~\cite{1999_Ducasse}, minimal clone length and maximal edit distance
 were doubled to 20 and 10, respectively.  Generated code that is
not subject to manual editing was excluded from clone detection, since
inconsistent manual updates obviously cannot occur. Normalization of
identifiers and constants was tailored as appropriate for the analyzed
language, to allow for renaming of identifiers while at the same time
avoiding too large false positive rates. These settings were
determined to represent the best compromise between precision and
recall during cursory experiments on the analyzed systems, for which
random samples of the detected clones have been evaluated manually.

The detected clone candidates were then manually rated by the researcher in
order to remove false positives, \ie code fragments that, although identified
as clone candidates by the detection algorithm, have no semantic relationship.
Inconsistent and exact clone group candidates were treated differently: \emph{all}
inconsistent clone group candidates were rated, producing the set of
inconsistent clone groups. Since the exact clones were not required for
further steps of the case study, instead of rating all of them, a random sample
of 25\% was rated, and false positive rates then extrapolated to determine the
number of exact clones.

\begin{figure*}
\centering
\includegraphics[scale=.54]{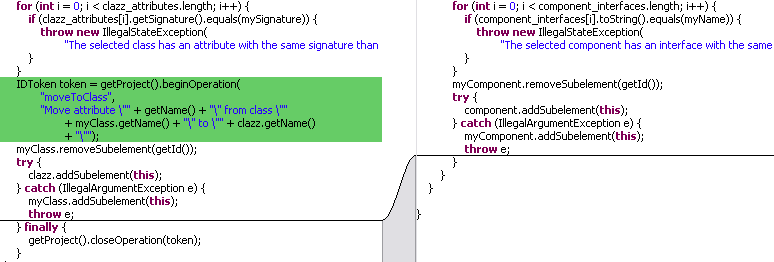} 
\caption{Different UI behavior since right side does not use operations (Sysiphus).}
\label{f:clone_ui}
\end{figure*}

The inconsistent clone groups were then presented to the developers
of the respective systems in the tool \emph{CloneDetective} mentioned
in Section~\ref{sec:tools}, which is able to display the commonalities
and differences of the clone group in a clearly arranged way, as
depicted in Figs.~\ref{f:clone_npe} and~\ref{f:clone_ui}. The
developers rated whether the clone groups were created intentionally
or unintentionally. If a clone group was created unintentionally, the
developers also classified it as faulty or non-faulty. For the Java
and C\# systems, all inconsistent clone groups were rated by the
developers. For the Cobol system, rating was limited to a random
sample of 68 out of the 151 inconsistent clone groups, since the age
of the system and the fact that the original developers were not
available for rating increased rating effort. Thus, for the Cobol
case, the results for RQ 2 and RQ 3 were computed based on this
sample. In cases where intentionality or faultiness could not be
determined, \eg because none of the original developers could be
accessed for rating, the inconsistencies were treated as intentional
and non-faulty.

\Section{Results}

The quantitative results of our study are summarized in
Table~\ref{tab:results}. Except for the Cobol system D, the precision
values are smaller for inconsistent clone groups than for exact clone
groups, as was expected, since inconsistent clone groups allow for
more deviation. The high precision results of system D result from the
rather conservative clone detection parameters chosen due to the
verbosity of Cobol. For system A, stereotype database access code of
semantically unrelated objects gave rise to lower precision values.

About half of the clones (52\%) contain inconsistencies.
Therefore, RQ 1 can be positively answered: Clones are changed inconsistently.
All these would not be reported by existing tools that search for exact
matches. From these inconsistencies over a quarter (28\%) has been introduced
unintentionally. Hence, RQ 2 can also be answered positively: Inconsistent
clones are created unintentionally in many cases. Only system D is far lower
here, with only 10\% of unintentionally inconsistent clones.  With about three
quarters of intentional changes, this shows that cloning and changing code
seems to be a frequent pattern during development and maintenance.

For RQ 3, whether inconsistent clones are indicators for
faults, we note that at least 3-23\% of the inconsistencies actually presented
a fault. Again the by far lowest number comes from the Cobol system.
Ignoring it, the total ratio of faulty inconsistent clones goes up to 18\%.
This constitutes a significant share that needs consideration.
To judge hypothesis H, we also calculated the fault densities. They
lie in the range of 3.4--91.4 faults per kLOC. Again, system D is an outlier.
Compared to reported fault densities in the range of 0.1 to
50 faults and considering the fact that all systems are not only
delivered but even have been productive for several years we consider
our results to support hypothesis H. On average the inconsistencies
contain more faults than average code. Hence, RQ 3 can also be answered
positively: Inconsistent clones can be indicators for faults in real
systems.

While the numbers are similar for the C\# and Java projects, rates of
unintentional inconsistencies and thus faults are comparatively low for project
D, which is a legacy system written in Cobol. To a certain degree, we attribute
this to our conservative assessment strategy of treating inconsistencies whose
intentionality and faultiness could not be unambiguously determined as
intentional and non-faulty. Furthermore, interviewing the current maintainers
of the systems revealed that cloning is such a common pattern in Cobol systems,
that searching for duplicates of a piece of code is actually an integral part
of their maintenance process. Compared to the developers of the other projects,
the Cobol developers where thus more aware of clones in the system. To account
for this difference in ``clone awareness'' we added the row $|F|/|UIC|$ to
Table~\ref{tab:results}, which reveals that while the rates of unintentional
changes are lower for project D, the ratio of unintentional changes leading to
a fault is in the same range for all projects. From our results it seems that
about every second to third unintentional change to a clone leads to a fault.

Although not central to our research questions, the detection of faults almost
automatically raises the question for their severity. As the fault effect costs
are unknown for the analyzed systems, we cannot provide a full-fledged severity
classification. However, we provide a partial answer by categorizing the found
faults as (1) faults that lead to potential system crash or data loss, (2)
unexpected behavior visible to the end user and (3) unexpected behavior not
visible to the end user. One example for a category (1) fault is shown in
Fig~\ref{f:clone_npe}. Here, one clone of the affected clone group performs a
null-check to prevent a null-pointer dereference whereas the other does not.
Other examples we encountered for category (1) faults are index-out-of-bounds
exceptions, incorrect transaction handling and missing rollbacks.
Fig.~\ref{f:clone_ui} shows an example of a category (2) fault. In one clone the
performed operation is not encapsulated in an operation object and, hence,
is handled differently by the undo mechanism. Further examples we found for category (2) faults are
incorrect end user messages, inconsistent default values as well as different
editing and validation behavior in similar user forms and dialogs. Category
(3) examples we identified include unnecessary object creation, minor memory leaks,
performance issues like missing break statements in loops and
redundant re-computations of cache-able values,  differences in exception
handling, different  exception and debug messages or different log levels for
similar cases.
Of the 107 inconsistent clones found, 17 were categorized as category (1)
faults, 44 as category (2) faults and 46 as category (3) faults. Since all
analyzed systems are in production, the relatively larger amounts of category
(2) and (3) faults coincide with our expectations.

\Section{Threats to validity} \vspace{-2mm}

We discuss how we mitigated threats to construct, internal and external
validity of our study.

\SubSection{Construct validity}

We did not analyze the development repositories of the systems in
order to determine if the inconsistencies really have been
introduced by incomplete changes to the system and not by random
similarities of unrelated code. This has two reasons: (1) We want to
analyze all inconsistent clones, also the ones that have been
introduced directly by copy and modification in a single commit. Those
might not be visible in the repository.  (2) The industrial systems do
not have complete development histories. We confronted this threat by
manually analyzing each potential inconsistent clone.

The comparison with average fault probability is not perfect to determine
whether the inconsistencies are really more fault-prone than a random
piece of code. A comparison with the actual fault densities of the systems
or actual checks for faults in random code lines would better suit this
purpose. However, the actual fault densities are not available to us because
of incomplete defect databases. To check for faults in random code lines
is practically not possible. We would need the developers time and willingness
for inspecting random code. As the potential benefit for the developers is
low, the motivation would be low and hence the results would be unreliable.

\SubSection{Internal validity}

As we ask the developers for their expert opinion on whether an inconsistency
is intentional or unintentional and faulty or non-faulty, a threat is
that the developers do not judge this correctly. One case is that
the developer assesses something as non-faulty which actually is faulty.
This case only reduces the chances to positively answer the research questions.
The second case is that the developers rate something as faulty which is
no fault. We mitigated this threat by only rating an inconsistency as faulty
if the developer was completely sure. Otherwise it was postponed and the
developer consulted colleagues that know the corresponding part of the code
better. Inconclusive candidates were ranked as intentional and non-faulty.
Hence, again only the chance to answer the research question positively is
reduced.

The configuration of the clone detection tool has a strong influence on the
detection results. We calibrated the parameters based on a pre-study and our
experience with clone detection in general. The configuration also varies over the different
programming languages encountered, due to their differences in
features and language constructs. However, this should not strongly
affect the detection of inconsistent clones because we spent great care to
configure the tool in a way that the resulting clones are sensible.

We also pre-processed the inconsistent clones that we presented to the
developers in order to eliminate false positives. This could mean that we
excluded clones that are actually faulty. However, this again only reduces the
chances that we can answer our research question positively.

\SubSection{External validity} \vspace{-2mm}

The projects were obviously not sampled randomly from all possible software
systems but we relied on our connections with the developers of the systems.
Hence, the set of systems is not completely representative. The majority of the
systems is written in C\# and analyzing 5 systems in total is not a high number.
However, all 5 systems have been developed by different development organizations and
the C\#-systems are technically different (2 web, 1 rich client) and provide
substantially different functionalities. We further mitigated this threat by
also analyzing a legacy Cobol system as well as an open source Java system.

\Section{Discussion} \vspace{-2mm}

Even considering the threats to validity discussed above, the results of the
study show convincingly that clones can lead to faults in a system. The
inconsistencies between clones are often not justified by different
requirements but can be explained by developer mistakes.

We consider of special
value the analysis of the Sysiphus project. Because both Sysiphus and our
detection tools are open source, the whole analysis can completely be
replicated independently. We provide a web site with the necessary
information\footnote{\url{http://wwwbroy.in.tum.de/~ccsm/icse09/}}.

Having established the empirical results, the question remains of how to use
this information in order to reduce faults in software systems. The answer is
twofold: (1) prevention by less cloning and (2) tools that prevent
unintentionally inconsistent changes of clones. The fewer clones there are in
the system, the less likely it is to introduce faults by inconsistencies
between them. In order to increase developer awareness of clones, we have
integrated our clone detection tool into the Visual Studio development
environment\footnote {\url{http://www.codeplex.com/CloneDetectiveVS}}. At the
Munich Re Group, as a reaction on the clone results, clone detection is now
included in the nightly builds of all discussed projects. Furthermore, for
existing clones, there should be tool support that ensures that all changes
that are made to a clone are made in the full knowledge of its duplicates.
Tools such as CloneTracker~\cite{2007_Ekoko_tracking_clones} or
CReN~\cite{2007_Jablonski_cren} provide promising approaches.
However, both
approaches are not applicable to existing software that already contains
inconsistent clones. Due to their high fault potential, we consider the ability
to detect inconsistent clones an important feature of industrial-strength clone
detectors.

\Section{Conclusion} 

In this paper we provide strong evidence that inconsistent clones
constitute a major source of faults, which means that cloning can be a
substantial problem during development and maintenance unless special
care is taken to find and track existing clones and their evolution.
Our results suggest that nearly every second unintentionally
inconsistent change to a clone leads to a fault.  Furthermore, we
provide a scalable algorithm for finding such inconsistent clones as
well as suitable tool support for future experiments.

Future work on this topic will evolve in multiple directions. One obvious
development is the refinement of the algorithms and tools used. This includes
refined heuristics to speed up the clone search and perform automatic
assessment to discard obviously irrelevant clones. In addition, the usability
of the tools could be advanced further to make their use more efficient for
practical applications. Moreover, it will be interesting to compare different
detection parameter values, algorithms and tools according to their performance
and accuracy when finding inconsistent clones.

Additionally, while answering some questions, our data of course
raises a couple of new relevant questions. One is a more detailed
quantitative classification of defect types of the faults
found. Another question is whether those faults are also detected by
classical techniques such as dynamic testing. However, to answer these
questions the developers of the analyzed systems have to be
interviewed again.

The underlying major question is how studying cloning can help in
reducing the development and maintenance costs of software
systems. This paper takes a first step into this direction, but more
work needs to be done to develop a usable and economically sensible
methodology.

Coming back to the paper title, we found that code clones do matter. Our
result is, however, limited to the consequences of clones on program
correctness. Hence, we believe that the most important task of future work is
to investigate the impact of clones on software maintenance effort.

\paragraph{Acknowledgments~} The authors would like to thank the Munich Re
Group,
LV 1871 and the Sysiphus team for supporting this study as well as Magne
J{\o}rgensen for helpful comments on the empirical analysis. This work has
partially been supported by the German Federal Ministry of Education and
Research (BMBF) in the project QuaMoCo (01 IS 08023B).

\bibliographystyle{latex8}
\bibliography{icse09}

(c) 2009 IEEE. Personal use of this material is permitted. Permission from IEEE must be obtained for all other users, including reprinting/ republishing this material for advertising or promotional purposes, creating new collective works for resale or redistribution to servers or lists, or reuse of any copyrighted components of this work in other works.

\end{document}